\documentclass[sigconf, 9pt]{acmart}

\usepackage{subfig,graphicx} 
\AtBeginDocument{%
  \providecommand\BibTeX{{%
    \normalfont B\kern-0.5em{\scshape i\kern-0.25em b}\kern-0.8em\TeX}}}

\copyrightyear{2022}
\acmYear{2022}
\setcopyright{acmlicensed}\acmConference[LP-IoT'21]{1st ACM Workshop on No Power and Low Power Internet-of-Things }{January 31-February 4 2022}{New Orleans, LA, USA}
\acmBooktitle{1st ACM Workshop on No Power and Low Power Internet-of-Things (LP-IoT'21), January 31-February 4 2022, New Orleans, LA, USA}
\acmPrice{15.00}
\acmDOI{10.1145/3477085.3478991}
\acmISBN{978-1-4503-8702-6/22/01}

\begin{document}
\title{smol: Sensing Soil Moisture using LoRa}
\author{Daniel Kiv$^*$, Garvita Allabadi$^*$, Berkay Kaplan, Robin Kravets}
\email{{dkiv2, garvita4, berkayk2,rhk}@illinois.edu}
\thanks{$^*$ Equal contribution}
\affiliation{%
  \institution{University of Illinois at Urbana-Champaign}
}

\begin{abstract}

Technologies for environmental and agricultural monitoring are on the rise, however there is a lack of small, low-power, and low-cost sensing devices in the industry. One of these monitoring tools is a soil moisture sensor. Soil moisture has significant effects on crop health and yield, but commercial monitors are very expensive, require manual use, or constant attention. This calls for a simple and low cost solution based on a novel technology.

In this work we introduce \textbf{\textit{smol}}: Sensing \textbf{S}oil \textbf{Mo}isture using \textbf{L}oRa, a low-cost system to measure soil moisture using received signal strength indicator (RSSI) and transmission power. It is compact and can be deployed in the field to collect data automatically with little manual intervention. Our design is enabled by the phenomenon that soil moisture attenuates wireless signals, so the signal strength between a transmitter-receiver pair decreases. We exploit this physical property to determine the variation in soil moisture. We designed and tested our measurement-based prototype in both indoor and outdoor environments. With proper regression calibration, we show soil moisture can be predicted using LoRa parameters.

\end{abstract}



\maketitle

\section{Introduction}
Data-driven agriculture techniques like precision agriculture promise to cut input costs (e.g. reduce fertilizer requirement), improve yields, and improve sustainability in agriculture \cite{vasisht2017farmbeats, Godfray812}. These techniques rely on accurate estimates of soil and plant properties, e.g. soil moisture, soil pH, crop density, leaf wetness, etc. using specialized sensors. Such sensors cost between hundreds and thousands of dollars each and become expensive to deploy at scale.\footnote{Some hobbyist sensors cost tens of  dollars each, but they are typically not used as agricultural sensors due to their limited accuracy.}

To avoid the cost of sensors, there have been multiple attempts at using radio signals to infer soil properties like soil moisture. Essentially, radio signals from already deployed sensors (or just wireless transmitters) are used to infer additional soil properties, reducing the cost of sensing. As RF signals travel through soil, their properties, like attenuation and phase, change as a function of soil properties. Therefore, past work~\cite{ding2019towards,wang2020soil} has shown that it is possible to measure soil moisture and electrical conductivity by measuring the characteristics of the wireless signal itself.

Past work has used Wi-Fi~\cite{ding2019towards} and  RFIDs~\cite{wang2020soil} for such analysis. We argue that such a design is infeasible for large scale deployments. First, Wi-Fi and RFIDs have limited range and, therefore, are not typically used in farm networks or for connecting farm sensors. Second, these designs rely on access to phase information of the signal which requires specialized hardware. Few Wi-Fi devices make this information available to application developers and RFIDs need specialized hardware to read the tags, let alone read phase values.

In this paper, we investigate the use of LoRa, a low-power long-range communication protocol, for sensing soil moisture. LoRa, given its low-power and long-range capabilities, is emerging as the front-runner for connecting Internet-of-things devices in outdoor environments \cite{bor2016lora}. Therefore, we do not need to deploy any additional hardware and can use devices with LoRa connectivity already present on the farms. Furthermore, we base our design to the use of RSSI (Received Signal Strength Indicator) parameters reported by commercial LoRa chipsets present in both gateways and end devices. 

We rely on LoRa devices already deployed in the farm and embedded in the soil. A LoRa receiver outside the soil measures the signal from such devices and estimates the signal strength. Intuitively, our design relies on the physical principle that wireless signals become progressively weaker as they traverse moist soil. The signal strength loss is more for moist soil as opposed to dry soil. Therefore, we should be able to use RSSI to identify soil moisture levels. The use of LoRa and RSSI offer us several advantages -- LoRa is low power, so sensors do not need frequent battery replacements. LoRa is ubiquitous, so our design does not need additional hardware. Finally, RSSI is an easy to use metric and available on multiple devices.

However, as compared to past work, restricting our measurements to signal strength indicators reduces the amount of available information. The phase values are more informative and have been used in most recent papers on wireless sensing for this reason~\cite{spotfi, chronos, bansal2021, gjengset2014}. To make up for the lack of phase information, we rely on computational tools and multiple measurements. Specifically, LoRa devices can transmit at multiple power levels. Low power levels support longer battery life, while the higher ones are needed for long range connections. We leverage this ability to perform multiple measurements of the signal strength at different power levels. We, then, use regression techniques to identify the moisture value from the multiple signal strength measurements.

In this paper, we make the following contributions:
\begin{itemize}
    \item We quantify the relationship between soil moisture and signal strength using a cheap, off-the-shelf LoRa chipset.
    \item We evaluate multiple setups (indoors vs outdoors, different heights) for measuring soil moisture based on LoRa signal strength.
    \item We leverage multiple transmit powers of LoRa devices to improve soil moisture estimation.
\end{itemize}
We implemented our design using the Adafruit M0 RFM9x 915MHz chipsets, which supports LoRa, and evaluated our design in multiple soil types and experimental settings -- soil in pots (indoors), farms (outdoors), etc. Our results demonstrate:
\begin{itemize}
    \item LoRa received signal strength correlates with volumetric water content, as measured by our ground truth.
    \item Machine learning regression models can map RSSI and transmission power to soil moisture. 
\end{itemize}

We envision that future iterations of our design will integrate with robots and drones to perform continuous soil moisture measurements. We also aim to leverage the motion of these drones or robots to improve our soil moisture estimates.

\section{Related Work}

Different metrics for measuring soil health include electrical conductivity. In recent years, wireless signals have been used in various sensing applications. In environmental and agricultural applications, wireless signals can be used to examine soil health with the basic premise that signal attenuation varies with different media \cite{ding2019towards, wan2017lora}. However, specialized devices to measure these metrics, that are currently employed commercially, are relatively expensive (sub-1000 USD) \cite{ding2019towards}.

WiFi time-of-flight (ToF) can be used to estimate electrical conductivity and soil moisture~ \cite{ding2019towards}. The bandwidth problem arises mainly from the system using time-of-flight as it requires high bandwidth from 100s of Mhz to few Ghz \cite{ding2019towards}. Strobe maps the propagation time and amplitude of WiFi signals received by different antennas to electrical conductivity, which depends on the soil moisture property \cite{ding2019towards}. Using this method, strobe can accurately estimate the soil moisture up to an error margin of 15\% \cite{ding2019towards}. A limitation is that to achieve good performance, ToF systems require ultra-wide bandwidth, which needs specialized equipment that will ramp up the cost to thousands of dollars \cite{ding2019towards}.

Previous research has also explored the relationship between RSSI and moisture level in the passive RFID domain. GreenTag~\cite{wang2020soil} explores the relationship between RSSI and soil moisture by using the reflection of the soil embedded tags' Differential Minimum Response Threshold (DMRT) value \cite{wang2020soil}. Researchers embed every tag into a plant pot, and the tags use a low-pass filtered DMRT to show that the DMRT is subject to changes in radio frequency environments and in the pot locations as water is added \cite{wang2020soil}. However, the shortcoming of this method is that the RFID reader is customized and expensive as like the previous example \cite{wang2020soil}.



Previous research using transmission power to aid in sensing is limited. One study leveraged transmission power, along with RSSI, in order to increase localization accuracy for wireless nodes \cite{bor2017}. Along a similar vein of RF-based localization, another study created \textit{Powerscan} which switches through different transmission powers and records different RSSI values \cite{wolfgang2016L}. The authors noticed a fixed RSSI offset between transmission powers had some saturation at low distances, which led to their decision to select a near median transmission power. One study models wireless sensor networks and investigates wave propagation of underground networks by examining the effect of volumetric water content on power density. They find that soil moisture attenuates wireless signals and decreases the power. However, their implementation doesn't examine the affects in an underground to above-ground environment \cite{dong2011}. Finally, one study investigates LoRa link quality for underground to above-ground transmissions, but the signal strength was not mapped to soil moisture \cite{lin2019underground}.

While there is some work mapping RSSI to another feature using models, such as machine learning\cite{anjum2020, Daramouskas2019}, none blend the elements of LoRa, soil moisture, and transmission power. In this paper, we leverage LoRa and simple easy to access metrics using a variety of models to attain an estimate of soil moisture.
\section{Design}

\begin{figure}\centering
\subfloat[Training Mode]{\label{softwaresetup-a}\includegraphics[width=.43\linewidth,height=7cm]{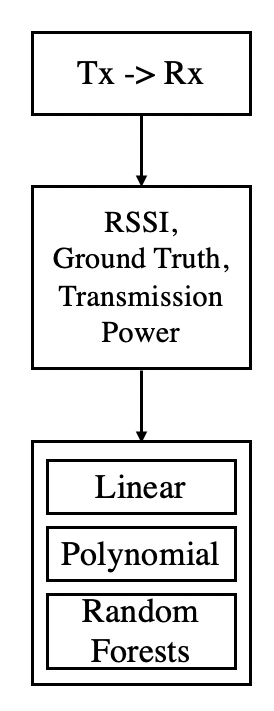}}\hfill
\subfloat[Inference Mode]{\label{softwaresetup-b}\includegraphics[width=.43\linewidth,height=7cm]{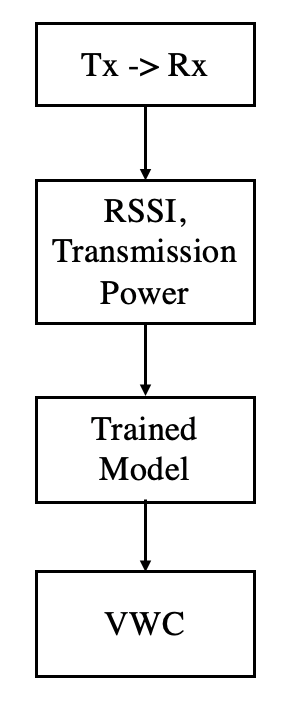}}\hfill 
\vspace{-0.1in}
\caption{smol Software Design: In the training phase, the receiver measures the RSSI of the signal and the ground truth. smol uses this data to train a regression model. In the test phase, the trained model predicts Volumetric Water Content (VWC) from RSSI measurements.} 
\vspace{-0.1in}
\label{softwaresetup}
\end{figure}

\textit{smol} is a LoRa based soil moisture sensing system. It uses two LoRa transceivers. The transmitter device is placed inside the soil, while the receiver can be at varying heights from the soil bed, in air.  A Raspberry Pi hosts the communication and data collection for the devices.  Figure \ref{drone} shows a depiction for one of the uses case of smol. The transmitter sweeps through the different power levels on the device and broadcasts the power information to the receiver. The receiver records the Received Signal Strength and the power information. Once the data is collected, smol leverages this data for soil moisture estimation using regression models. Figure \ref{softwaresetup} describes the two modes of operation for smol. In the training mode, the receiver combines the measured values with the ground truth supplied by an industry scale moisture sensor to train regression models. While in the test mode, it uses this information to predict the Volumetric Water Content (VWC) of the soil using the trained regression model. 

We discuss the details of the key features and design choices for smol in the subsequent sections. 

\subsection{Long Range (LoRa)}

LoRa uses Chirp Spread Spectrum (CSS) modulation, a spread spectrum technique where the wireless signal is modulated by chirp pulses (frequency varying sinusoidal pulses) hence improving resilience and robustness against the Doppler and multi-path effect \cite{adelantado2017understanding}. Previous work identifies LoRa as more stable compared to WiFi or BLE in terms of higher adaptability to the changes in the environment \cite{islam2019lorain}. However, LoRa is not resistant to the variables that could affect the signal from an underground node to an above-ground node. Some of these include attenuation, reflection, and multi-path fading \cite{vuran2009}.

The radio can be modulated by a variety of parameters such as transmission power, spread factor, and bandwidth \cite{lin2019underground}. Spread factor determines the number of chirps per second. A higher spread factor increases the sensitivity and energy cost. Bandwidth represents the frequency range of the signals. A higher bandwidth yields a higher data rate but also at the cost of distance. Transmission power affects the communication performance and range of the signal at the cost of energy. That is, a higher transmission power yields a lower signal-to-noise ratio and higher range but also requires more energy. At the opposite end, we can save energy but have a lower range. By sweeping across the full range of transmission powers, multiple measurements can be taken for some soil moisture level. Instead of selecting a single transmission power, such as the highest one, a balance between the two extremes can be reached.

LoRa typically operates in the sub-GHz frequency band making it ideal for long range transmissions. A higher frequency band also ensures that LoRa has deeper soil penetration. All these factors make LoRa a good choice for using in a soil moisture sensing system.

\subsection{Ground Truth Measurements}

Typically soil is composed of soil, water and, air. The Volumetric Water Content (VWC) of soil is the ratio of volume of water in a given volume of soil to the total soil volume. This value is represented as a decimal or percentage. 

For our ground truth values, we use an industry scale soil moisture sensor - FieldScout TDR 300 Soil Moisture Meter \cite{fieldscout}. The sensor uses Time Domain Reflectometry (TDR) to measure the VWC of soil. The underlying principal of TDR involves measuring the travel time of an electromagnetic wave along a wave-guide \cite{fieldscoutmanual,josephson2019}. The speed of the wave in soil is dependent on the relative permittivity of the soil. Recall that  relative permittivity or the dielectric permittivity constant ($\varepsilon_\mathrm{r}$), is the ratio of the permittivity of the substance to the permittivity of free space, $\varepsilon_0$.

\begin{equation} \label{permittivity}
    \varepsilon_\mathrm{r} = \frac{\varepsilon}{\varepsilon_0}
\end{equation}

The fact that water ($\varepsilon_\mathrm{r}$ = 80) has a much greater dielectric constant than air ($\varepsilon_\mathrm{r}$ = 1) or soil solids ($\varepsilon_\mathrm{r}$ = 3-7) is exploited to determine the VWC of the soil. The VWC measured by TDR is an average over the length of the waveguide. 

We make two observations here. First, the ground truth sensor relies on the same principle which guides wireless signal attenuation. Wireless signal attenuation depends on the electric permittivity of the medium -- in this case, soil. As soil moisture increases, the electric permittivity increases and the signal strength decreases for typical wireless transmissions. Therefore, we are relying on a proxy measure for our system similar to traditional sensor. Second, the ground truth sensor costs approximately 1500 US dollars per sensor for errors of $\pm 3\%$.

\subsection{Calibration using regression models}

\begin{figure}\centering
\subfloat[RSSI decreases with volumetric water content. Air (bar on the left) and water (bar on the right) are provided as baselines. The ground truth was calibrated against tap water.]{\label{a}\includegraphics[width=\linewidth,height=7cm,keepaspectratio]{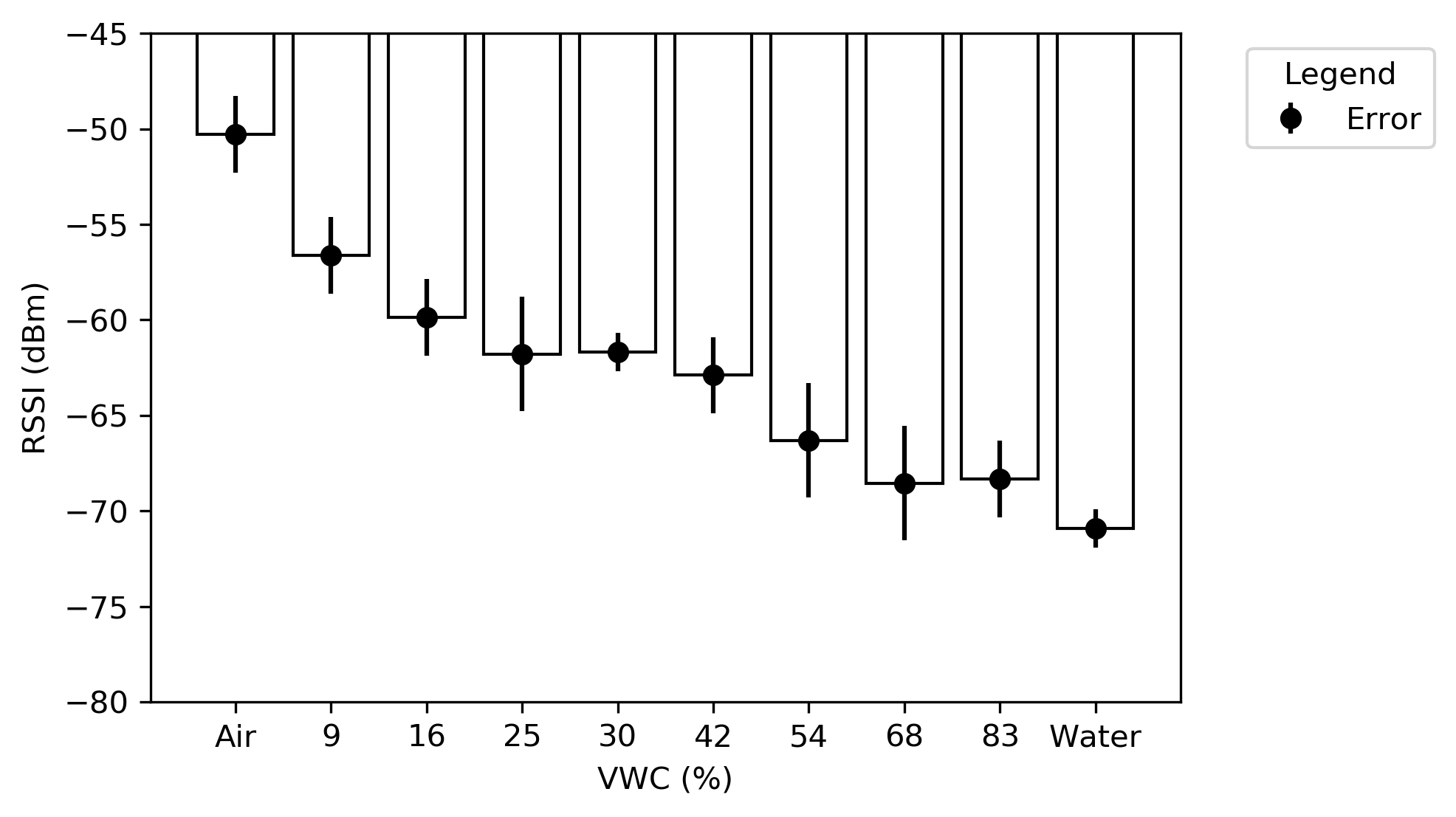}}\hfill
\subfloat[All possible transmission power supported by the transceiver.]{\label{b}\includegraphics[width=\linewidth,height=7cm,keepaspectratio]{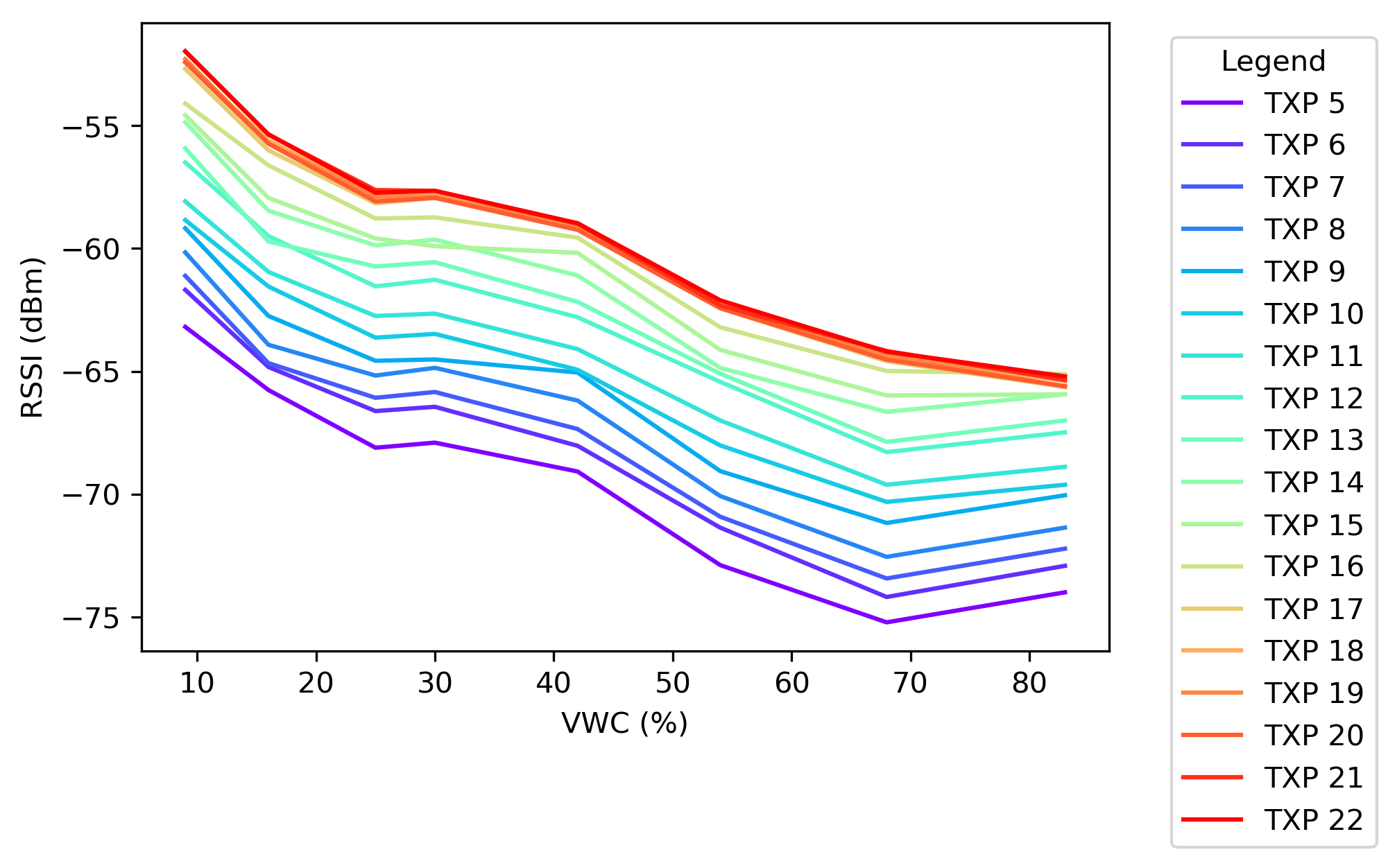}}\hfill 
\vspace{-0.1in}
\caption{Initial indoor bucket tests with the receiver on top of the soil and the transmitter buried 15 cm.} 
\vspace{-0.1in}
\label{moisture}
\end{figure}

In order to map the features of the transceiver to soil moisture, the devices need to be in the communication range to exchange packets. LoRa works fairly long range compared to other communication devices, even in an underground setting \cite{wan2017lora}. However, most of the experiments we perform will have a distance of less than 300 cm between the two transceivers. After setting up the distance, we calibrate the ground truth instrument. The 0\% value is calibrated to air while 100\%, full moisture, is calibrated to water.

Once the ground truth is calibrated, we begin the calibration process for the LoRa transceivers. In the experiments, a transmitter is buried 15 cm in the soil and the receiver is vertically placed at a height at or above the soil. At each level of moisture, we incrementally increase the VWC by moderately adding water. The ground truth reading is recorded by taking an average of 10 measurements in various spots around the area where the transmitter was buried. Subsequently, a Python script runs to collect data at each step. During the process, the transmitter sweeps through the 18 available power levels, ranging from 5 to 23 dBm, broadcasting a single packet at each level. Each packet contains information about the transmission power. One can also embed a unique identifier into the packet for multiple devices. The receiver receives the packet and records the transmission power level and RSSI.

After collecting the data, a plot will look similar to Figure \ref{results-inside} where there is a visible trend between RSSI and VWC. It is evident that the soil moisture is within the baseline calibration we set for the ground truth.

smol then fits a model with the RSSI and transmission power as inputs in order to predict the VWC. The models are trained using 80\% of the dataset and evaluated using the remaining 20\% of the dataset. A variety of regression models are used including linear regression, ridge regression, and random forest. Performances are compared in the results section. To make things more accessible and easy for the end user, smol implementation uses Python and the scikit-learn library \cite{scikit-learn} for calibration. Machine learning and regression is used in a variety of calibration situations \cite{wijeratne2020, wang2020soil}. Once the best fit curve is constructed and the best algorithm is chosen based on the metrics, RSSI and transmission can be continuously mapped to VWC.
\section{Implementation}

We implement smol  using two Adafruit M0 RFM9x 915MHz LoRa transceivers that use RSSI as the main metric \cite{adafruit}. One device acts as a transmitter while the other acts as a receiver. We also added a 7.5cm antenna to each of the devices to increase the range. The choice of hardware for our experiments was based on the available radio frequency bands, range of transmission powers, ease of programming and finally the cost of the device. Apart from these devices, smol also uses a Raspberry Pi 4 for communicating with the LoRa devices and collecting the data. 

smol is an easy to use, low-cost system to measure the soil moisture content. It uses a variety of machine learning models like linear regression, ridge regression, and random forests to train and predict the moisture content of soil. For this purpose, it measures three variables:

\begin{itemize}
    \item \textbf{Received signal strength indicator (RSSI).} The primary quantity of interest for correlating with soil moisture.
    \item \textbf{Transmission power from the transmitter.} The lower the transmission power, the lower the signal strength and possible range. The higher the transmission power, the higher the signal strength and possible range. This value can vary from 5 to 23 dBm. In the results, we mainly look at 5 to 22, since 23 dBm appears to wrap around to 5. This parameter can be easily modified with the provided Adafruit libraries.
    \item\textbf{Volumetric water content.} This is the ground truth value as measured by an industry scale soil moisture sensor.
\end{itemize}

\section{Experimental Setup}

The goal of our experiments is to examine how RSSI is affected by different levels of soil moisture and the distance between the sensors. We evaluate our system in two setups - inside the laboratory and in a real world scenario outside on a farm. 

\begin{figure}\centering
\subfloat[Vertical setup in the lab.]{\label{ex-a}\includegraphics[width=.3\linewidth,height=4cm]{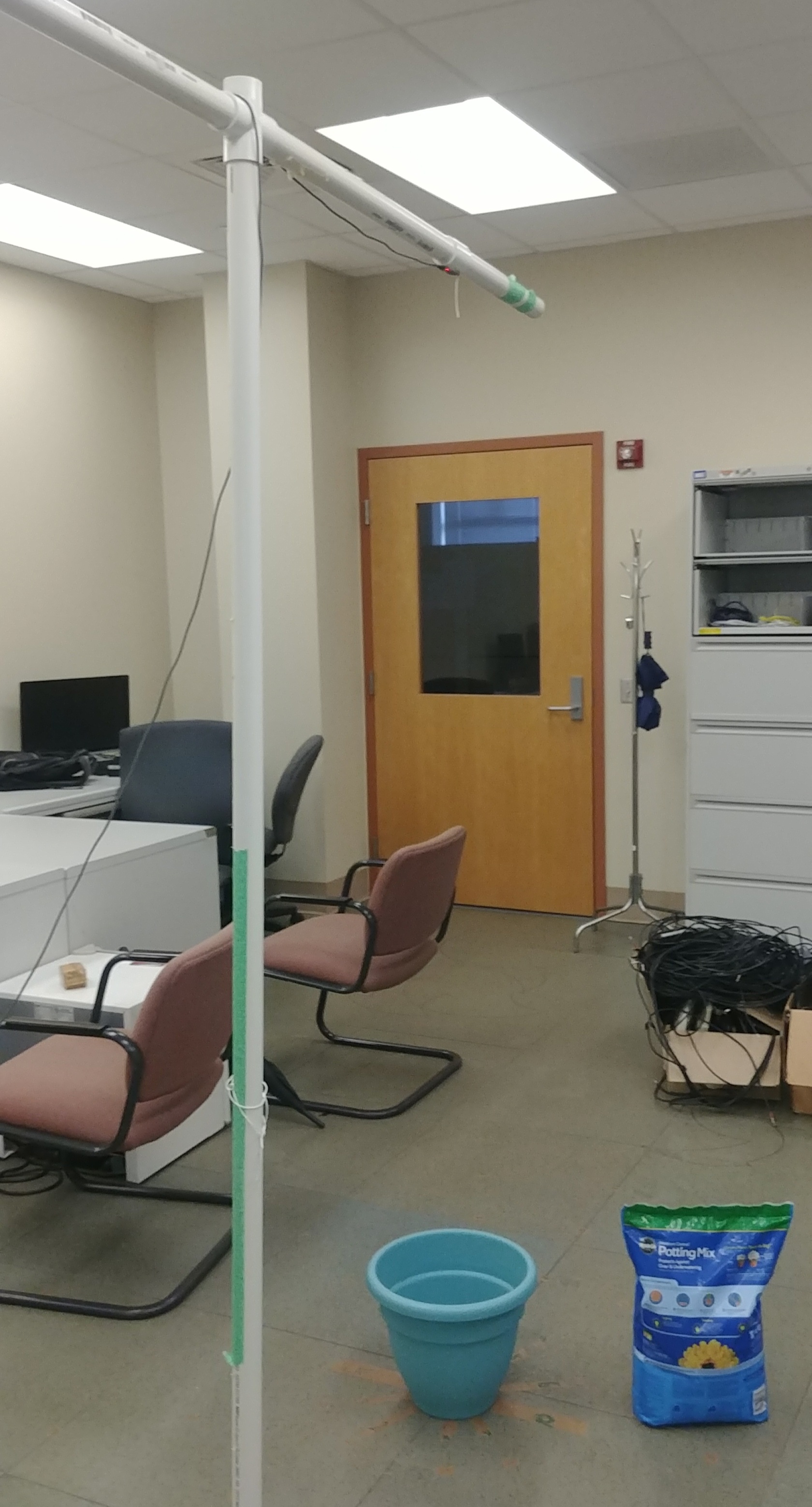}}\hfill
\subfloat[Vertical setup at research field.]{\label{ex-b}\includegraphics[width=.3\linewidth,height=4cm]{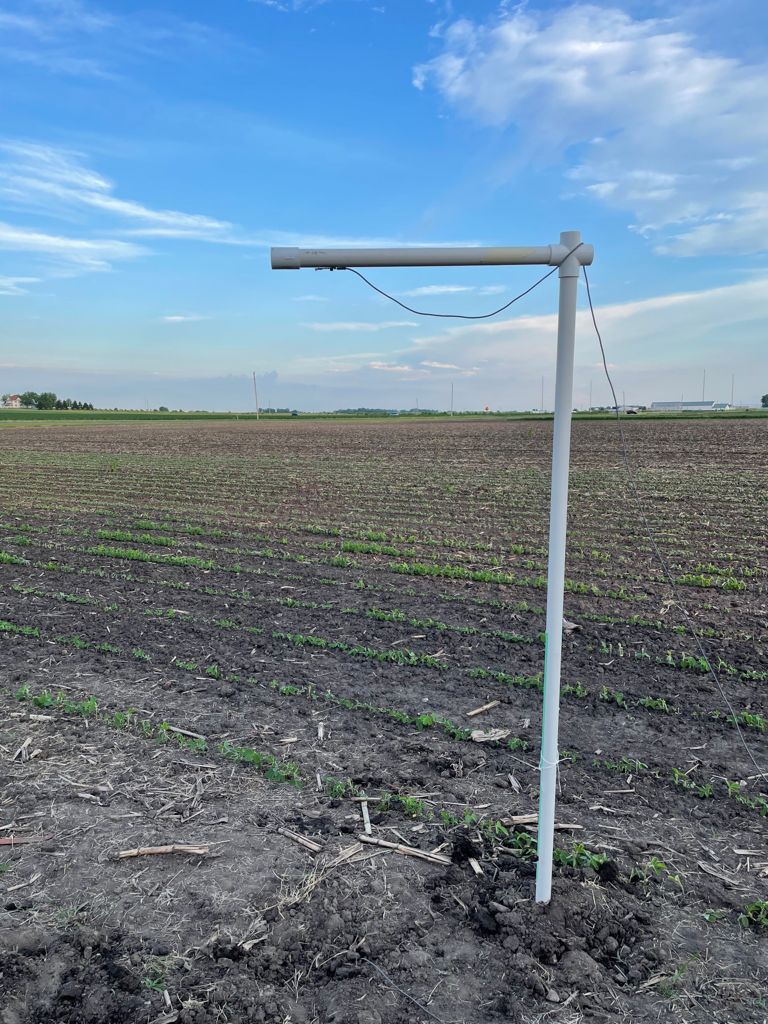}}\hfill 
\subfloat[Transmitter placed 15 cm in a hole.]{\label{ex-c}\includegraphics[width=.3\linewidth,height=4cm]{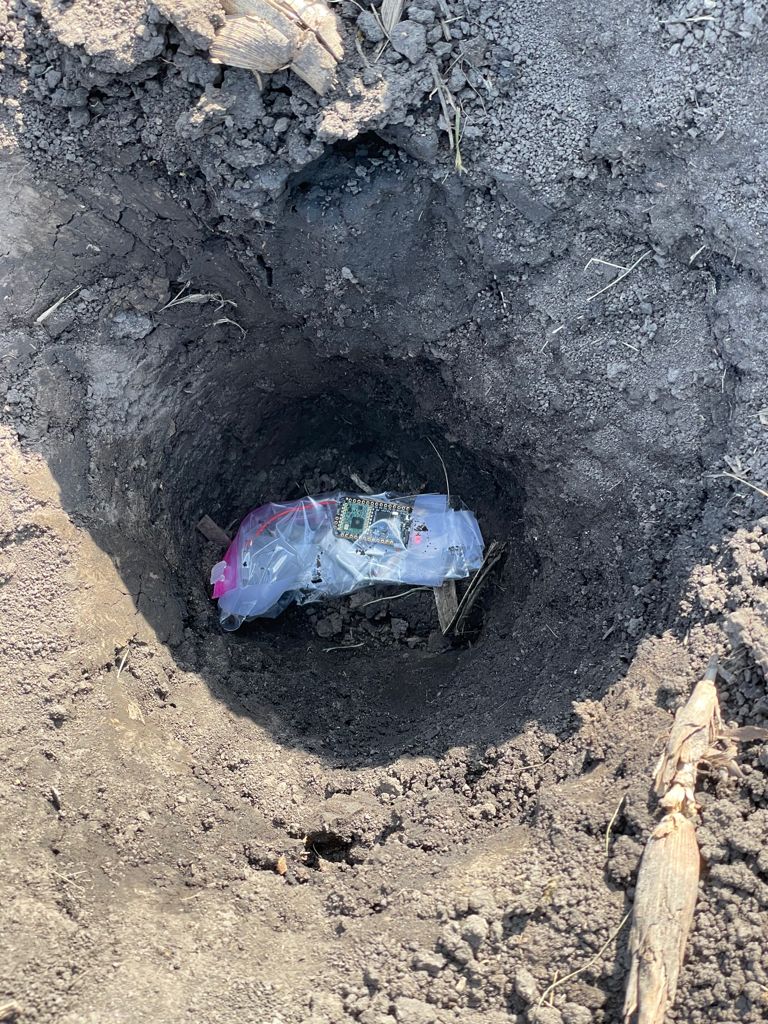}}
\caption{Experimental setups}
\label{ex-setup}
\end{figure}

\begin{figure}[t]
  \centering
  \includegraphics[width=\linewidth, height=8cm,keepaspectratio]{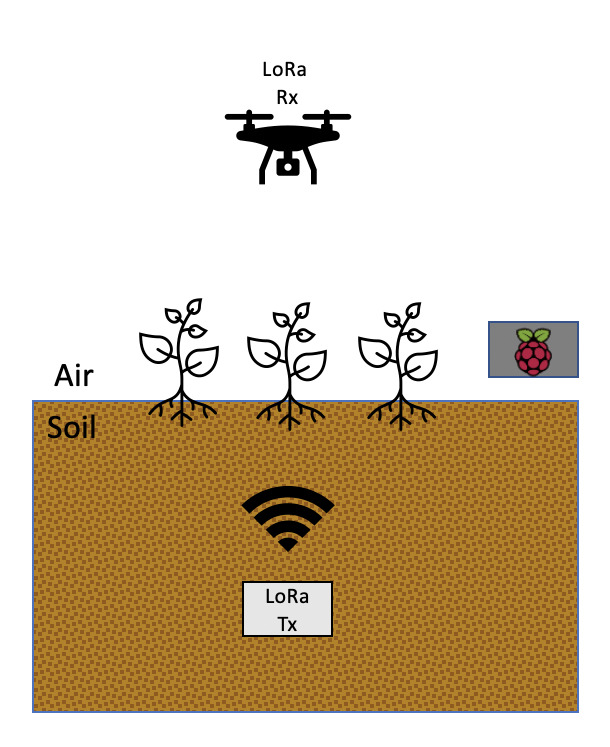}
  \vspace{-0.1in}
  \caption{An illustration of a possible extension for \textit{smol}, where a drone could take measurements based on the wireless signal strength from LoRa. This requires considering the movement of the drone, multi-path, obstructions, and other factors. 
  }
  \label{drone}
  \vspace{-0.1in}
\end{figure}

In the lab experiment, the transmitter was placed at the bottom of the bucket, 15cm in the soil. The receiver was placed at heights 0, 195 and 265cm away from the soil. It was attached to the arm of the PVC pipe so that it would be directly above the transmitter. The setup is depicted in Figure \ref{ex-a}. In this way, it could emulate the way that a drone or vehicles passes over the area to take a measurement, similar to Figure \ref{drone}. The devices were faced towards each other. The soil used in the lab was Miracle Gro Potting Mix \cite{miraclegro}. 

At the farms, we dug a hole approximately 15cm in an area at the plot. We secured the PVC pipes into the ground as shown in Figure \ref{ex-b}. The transmitter was placed inside the hole and then buried. This was done at the Illinois Autonomous Farm \cite{autofarm}. 

The transmitter was wrapped inside a plastic sandwich bag. The bag was twisted and taped in order to prevent water from damaging the electronics. Using a plastic bag is a low-cost way to protect electronics.\footnote{No devices were harmed in the making of this project.} Special care was taken so that the front facing part of the device had a single layer of the plastic bag in the line of sight between the transceivers. A visualization of the experimental setup can be seen in Figure \ref{ex-c}.

The soil was artificially moistened with water. This was done by sprinkling about 16 ounces of water in the soil. A few minutes were given to let the water settle, so that all the moisture was not at the top.

\section{Results}

\begin{figure}\centering
\subfloat[Laboratory measurements.]{\label{res-a}\includegraphics[width=\linewidth,height=8cm,keepaspectratio]{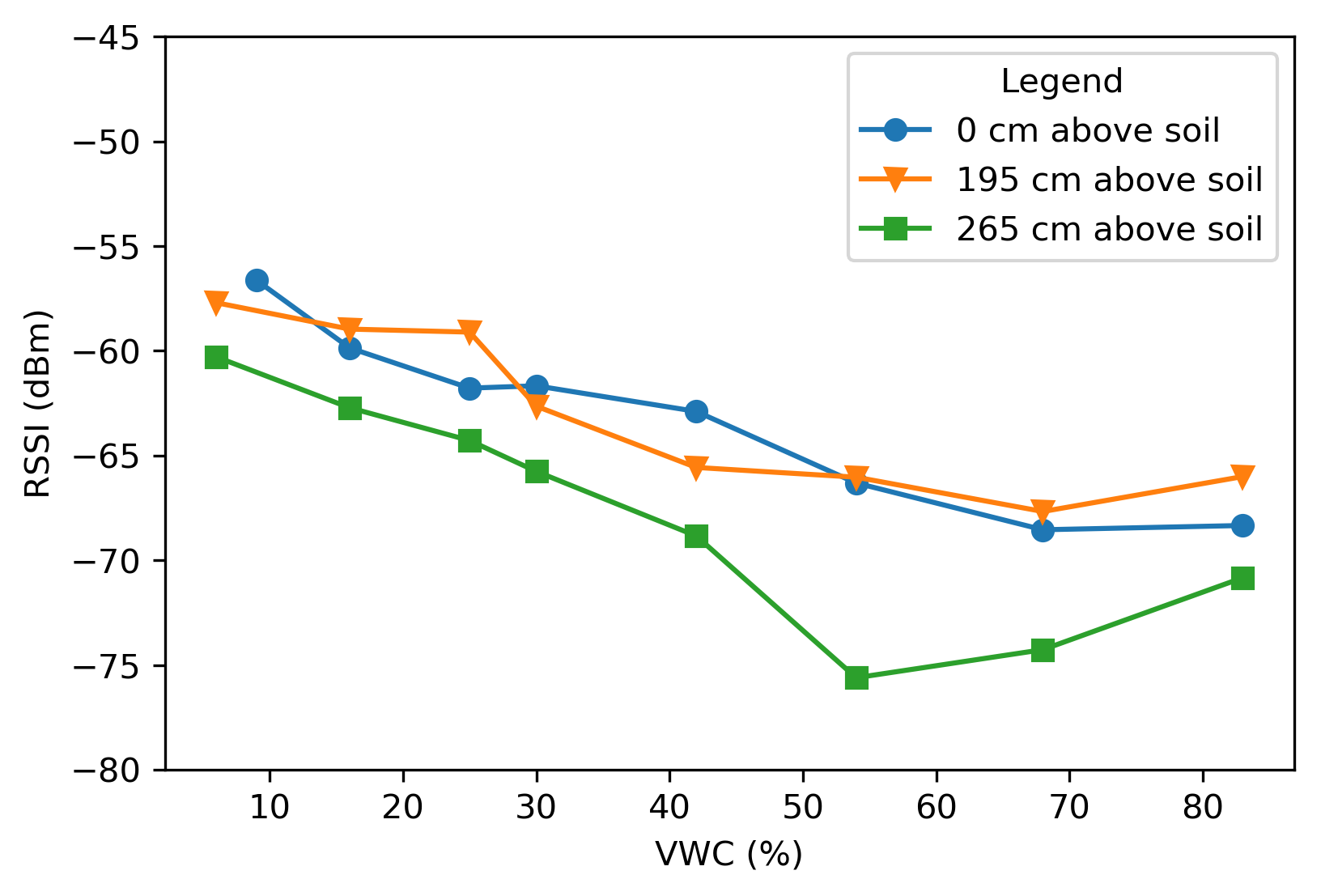}}\hfill
\subfloat[In-situ measurements at the research farm.]{\label{res-b}\includegraphics[width=\linewidth,height=8cm,keepaspectratio]{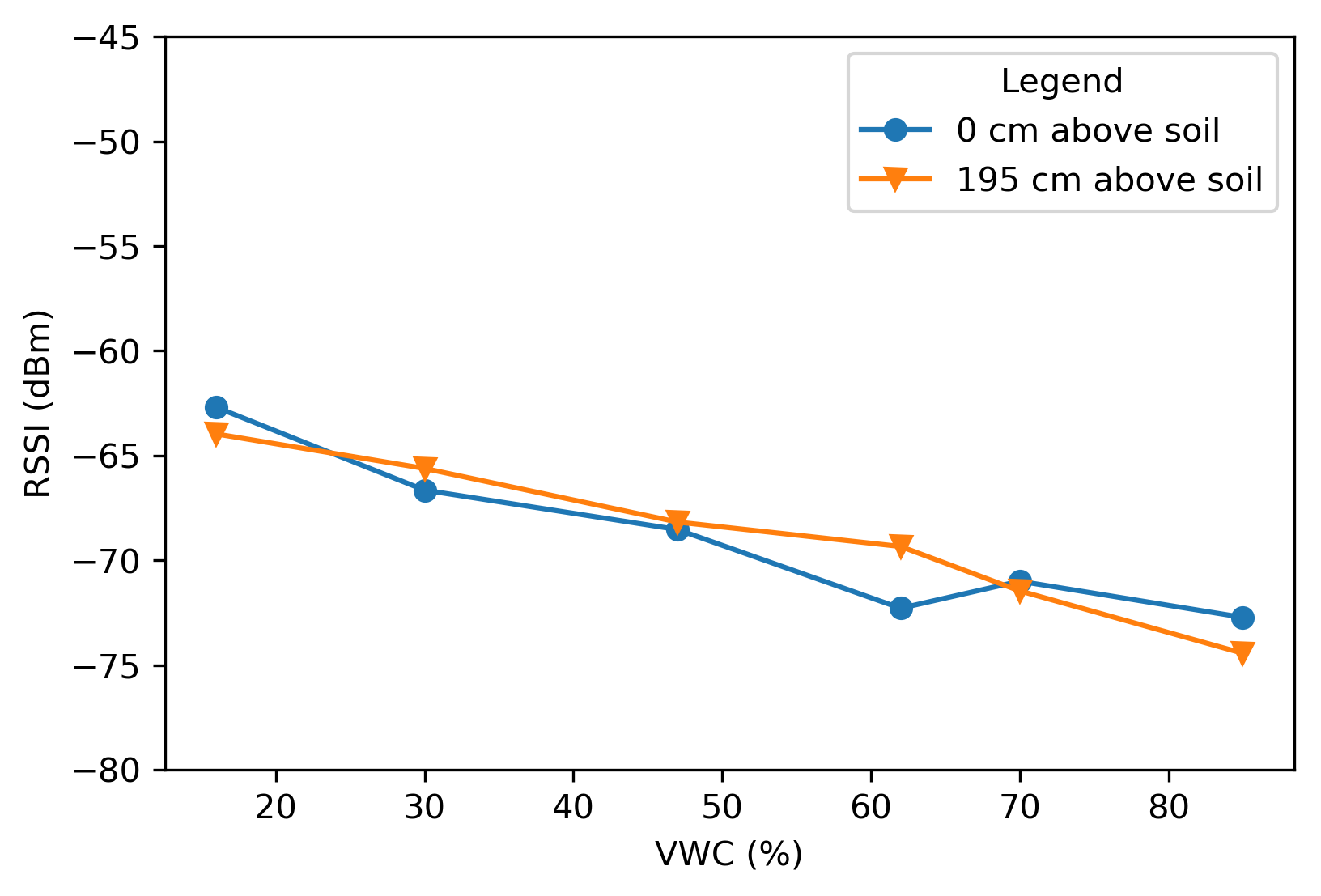}}\hfill 
\vspace{-0.1in}
\caption{Laboratory and in-situ results of our experiments. Each point on the graph represents the mean RSSI of the median transmission power. Each line is the height above the top of the soil that the recording occurred at.} 
\vspace{-0.1in}
\label{results-inside}
\end{figure}

\subsection{Laboratory and in-situ}

Figure \ref{results-inside} shows both the laboratory and in-situ results of our experiments. Each point on the graph represents the mean RSSI of the median transmission power. Each line is the height above the top of the soil that the recording occurred at. Since the offsets of the transmission powers, as shown in Figure \ref{moisture}, are generally equal, the overall curve would look similar. Selecting a different transmission power other than the median would just result in an offset.

Two baselines were obtained in air and only water as shown in Figure \ref{moisture}. As expected, with no obstructions in the line of sight, the signal strength was at its highest. In all water, the signal strength was at its lowest. Surprisingly, there was a decent signal strength even when completely submerged in water, which wireless signals tend to travel poorly through.

During the laboratory experiments, particularly for higher heights and VWC, the multi-path effect may affect RSSI readings. This is because the radius of the bottom of the pot, where the transmitter was placed, is smaller than the distance to the top of the soil from the bottom. The signal may take the path along the bottom radius and through the air, which could be faster than travelling straight through the soil. 

For comparison, we also performed the experiment \textit{in-situ} at the research farms of the University of Illinois at Urbana-Champaign. The multi-path effect would be reduced since the signals would have to take a route through the soil, since all around is dirt. The RSSI for the outdoor experiments for the same height are slightly lower as a result. For this experiment, we watered a larger area around where the transmitter was buried in order to better simulate rain in an open field.

\begin{table*}[h]
\centering
\begin{tabular}{|l|l|c|}
\hline
Model & $R^2$ & Mean Absolute Error (MAE) \\
\hline 
\textbf{Random Forest w/ all TX powers} & \textbf{0.92} & \textbf{1.63} \\
Random Forest w/ median TX power & 0.90 & 0.94 \\
\hline
Polynomial w/ all TX powers & 0.80 & 4.43 \\
Polynomial w/ median TX power & 0.88 & 1.45 \\
\hline
Linear Regression w/ all TX powers & 0.18 & 7.66 \\
Linear Regression w/ median TX power & 0.88 & 3.23 \\
\hline
\end{tabular}
\caption{Comparison of model performances.}\vspace{-0.2in}
\label{tab:models}
\end{table*}

\subsection{Regression models}

We evaluated and compared a variety of regression models. We use a random forest, polynomial with degree more than and equal to 2 (>= 2), and a linear regression. In the models appended with "all TX powers," the inputs are transmission power and RSSI. In models appended with "median TX power," the input is simply the RSSI from the median TX power. The output is the VWC. Basically, we are seeing how much of a performance gain we can get from adding an additional feature, the transmission power. Table \ref{tab:models} shows a comparison of the calibration models we trained and evaluated. Each model in the table provides the coefficient of determination ($R^2$) and mean absolute error (MAE). In general, higher $R^2$ value and a lower MAE value is better.

We can see that the random forest algorithm outperforms the other regression models, since it is better suited for multiple features. When comparing the two random forest models, we can see that having transmission power along with RSSI as an input can improve performance. Since transmission powers also scale with signal strength, devices don't need to be hard-coded with a single transmission power. Taking the highest transmission power is associated with higher energy cost. Taking the lowest transmission power causes the signal strength to be weak, which also means the maximum distance is shorter. By taking a spectrum, we can cover the entire range while also having comparable performance to just taking a single transmission power.

\section{Future Work and Limitations}

\textit{smol} is a first step to leverage the strengths of LoRa for agricultural monitoring. In this paper, we have shown experiments in a controlled setting, therefore there are some caveats that would require further investigation to make \textit{smol} more robust against factors that can affect the measurement.

An additional experiment can discuss the impact of different soil types such as sandy soil or clay soil etc. on the RSSI values for LoRa. It would be interesting to see the results of \textit{smol} for long scale continuous deployments in outdoor settings. One possible outcome for the system is to detect a rainy or dry day based on the signal strength. This can help in large scale deployment of \textit{smol}, making the system more easily accessible to those who have LoRa devices but are on a budget and want to deploy additional systems in their fields.

LoRa is known for long ranges. A study investigating how the signal strength varies at different soil depths, distances, and soil moisture would be interesting. Comparing the performance of measuring soil moisture using different wireless signals would be of great benefit. Each wireless technology has their own pros and cons. Depending on the situation, an optimal technology can be used to suit the user's intent.

Since LoRa signals were able to be received, by some extent, through water, it would be interesting to investigate the feasibility of using LoRa underwater. A work detailing the signal strength at different distances underwater could pave the way for a network that integrates with existing infrastructure, such as sound. Along the same vein, localization using LoRa signals underwater would be quite interesting.

Our setup calibrates each LoRa device that is used in the field. This may be a time consuming process, therefore work on improving the calibration process and generalizing to multiple devices would be of great benefit. It is important to note that the the regression model for the device was trained using data collecting in a controlled environment. There was no rain, foliage, or any significant obstruction between the devices. It is possible that rain could cause the air to become moist and shuffle the soil, which could affect the estimation of volumetric water content since the air could moist and occupy the space between the soil and the device. We leave this discussion for future work.

\section{Conclusion}

This paper presents smol, a low-power and low-cost LoRa based system to estimate the volumetric water content of soil. It leverages the received signal strength and transmission power information to estimate soil moisture. Our indoor and outdoor experiments demonstrate the correlation between RSSI and actual VWC. Results showed that \textit{smol} predicts the soil moisture levels accurately using trained regression models.

\section{Acknowledgements}

Thanks to Deepak Vasisht, Robin Kravets, and Vikram Adve for helpful discussions and providing hardware to support this work. We also want to thank the leadership at the Illinois Center for Digital Agriculture \cite{digitalag} for granting us access to the Illinois Autonomous Farm \cite{autofarm} to conduct our experiments. 

\bibliographystyle{ACM-Reference-Format}
\bibliography{sample-base}


\begin{thebibliography}{28}


\ifx \showCODEN    \undefined \def \showCODEN     #1{\unskip}     \fi
\ifx \showDOI      \undefined \def \showDOI       #1{#1}\fi
\ifx \showISBNx    \undefined \def \showISBNx     #1{\unskip}     \fi
\ifx \showISBNxiii \undefined \def \showISBNxiii  #1{\unskip}     \fi
\ifx \showISSN     \undefined \def \showISSN      #1{\unskip}     \fi
\ifx \showLCCN     \undefined \def \showLCCN      #1{\unskip}     \fi
\ifx \shownote     \undefined \def \shownote      #1{#1}          \fi
\ifx \showarticletitle \undefined \def \showarticletitle #1{#1}   \fi
\ifx \showURL      \undefined \def \showURL       {\relax}        \fi
\providecommand\bibfield[2]{#2}
\providecommand\bibinfo[2]{#2}
\providecommand\natexlab[1]{#1}
\providecommand\showeprint[2][]{arXiv:#2}

\bibitem[\protect\citeauthoryear{Adafruit}{Adafruit}{2016}]%
        {adafruit}
\bibfield{author}{\bibinfo{person}{Adafruit}.} \bibinfo{year}{2016}\natexlab{}.
\newblock \bibinfo{title}{Adafruit Feather M0 with RFM95 LoRa Radio - 900MHz -
  RadioFruit}.
\newblock \bibinfo{howpublished}{\url{https://www.adafruit.com/product/3178}}.
\newblock
\newblock
\shownote{Accessed on 04.30.2021.}


\bibitem[\protect\citeauthoryear{Adelantado, Vilajosana, Tuset-Peiro, Martinez,
  Melia-Segui, and Watteyne}{Adelantado et~al\mbox{.}}{2017}]%
        {adelantado2017understanding}
\bibfield{author}{\bibinfo{person}{Ferran Adelantado}, \bibinfo{person}{Xavier
  Vilajosana}, \bibinfo{person}{Pere Tuset-Peiro}, \bibinfo{person}{Borja
  Martinez}, \bibinfo{person}{Joan Melia-Segui}, {and} \bibinfo{person}{Thomas
  Watteyne}.} \bibinfo{year}{2017}\natexlab{}.
\newblock \showarticletitle{Understanding the limits of LoRaWAN}.
\newblock \bibinfo{journal}{\emph{IEEE Communications magazine}}
  \bibinfo{volume}{55}, \bibinfo{number}{9} (\bibinfo{year}{2017}),
  \bibinfo{pages}{34--40}.
\newblock


\bibitem[\protect\citeauthoryear{Anjum, Khan, Hassan, Mahmood, Qureshi, and
  Gidlund}{Anjum et~al\mbox{.}}{2020}]%
        {anjum2020}
\bibfield{author}{\bibinfo{person}{Mahnoor Anjum},
  \bibinfo{person}{Muhammad~Abdullah Khan}, \bibinfo{person}{Syed~Ali Hassan},
  \bibinfo{person}{Aamir Mahmood}, \bibinfo{person}{Hassaan~Khaliq Qureshi},
  {and} \bibinfo{person}{Mikael Gidlund}.} \bibinfo{year}{2020}\natexlab{}.
\newblock \showarticletitle{RSSI Fingerprinting-Based Localization Using
  Machine Learning in LoRa Networks}.
\newblock \bibinfo{journal}{\emph{IEEE Internet of Things Magazine}}
  \bibinfo{volume}{3}, \bibinfo{number}{4} (\bibinfo{year}{2020}),
  \bibinfo{pages}{53--59}.
\newblock
\urldef\tempurl%
\url{https://doi.org/10.1109/IOTM.0001.2000019}
\showDOI{\tempurl}


\bibitem[\protect\citeauthoryear{Bansal, Gadre, Singh, Rowe, Iannucci, and
  Kumar}{Bansal et~al\mbox{.}}{2021}]%
        {bansal2021}
\bibfield{author}{\bibinfo{person}{Atul Bansal}, \bibinfo{person}{Akshay
  Gadre}, \bibinfo{person}{Vaibhav Singh}, \bibinfo{person}{Anthony Rowe},
  \bibinfo{person}{Bob Iannucci}, {and} \bibinfo{person}{Swarun Kumar}.}
  \bibinfo{year}{2021}\natexlab{}.
\newblock \showarticletitle{OwLL: Accurate LoRa Localization Using the TV
  Whitespaces}. In \bibinfo{booktitle}{\emph{Proceedings of the 20th
  International Conference on Information Processing in Sensor Networks
  (Co-Located with CPS-IoT Week 2021)}} (Nashville, TN, USA)
  \emph{(\bibinfo{series}{IPSN '21})}. \bibinfo{publisher}{Association for
  Computing Machinery}, \bibinfo{address}{New York, NY, USA},
  \bibinfo{pages}{148–162}.
\newblock
\showISBNx{9781450380980}
\urldef\tempurl%
\url{https://doi.org/10.1145/3412382.3458263}
\showDOI{\tempurl}


\bibitem[\protect\citeauthoryear{Bor and Roedig}{Bor and Roedig}{2017}]%
        {bor2017}
\bibfield{author}{\bibinfo{person}{Martin Bor} {and} \bibinfo{person}{Utz
  Roedig}.} \bibinfo{year}{2017}\natexlab{}.
\newblock \showarticletitle{LoRa Transmission Parameter Selection}. In
  \bibinfo{booktitle}{\emph{2017 13th International Conference on Distributed
  Computing in Sensor Systems (DCOSS)}}. \bibinfo{pages}{27--34}.
\newblock
\urldef\tempurl%
\url{https://doi.org/10.1109/DCOSS.2017.10}
\showDOI{\tempurl}


\bibitem[\protect\citeauthoryear{Bor, Vidler, and Roedig}{Bor
  et~al\mbox{.}}{2016}]%
        {bor2016lora}
\bibfield{author}{\bibinfo{person}{Martin Bor}, \bibinfo{person}{John~Edward
  Vidler}, {and} \bibinfo{person}{Utz Roedig}.}
  \bibinfo{year}{2016}\natexlab{}.
\newblock \showarticletitle{LoRa for the Internet of Things}.
\newblock  (\bibinfo{year}{2016}).
\newblock


\bibitem[\protect\citeauthoryear{Daramouskas, Kapoulas, and
  Paraskevas}{Daramouskas et~al\mbox{.}}{2019}]%
        {Daramouskas2019}
\bibfield{author}{\bibinfo{person}{Ioannis Daramouskas},
  \bibinfo{person}{Vaggelis Kapoulas}, {and} \bibinfo{person}{Michael
  Paraskevas}.} \bibinfo{year}{2019}\natexlab{}.
\newblock \showarticletitle{Using Neural Networks for RSSI Location Estimation
  in LoRa Networks}. In \bibinfo{booktitle}{\emph{2019 10th International
  Conference on Information, Intelligence, Systems and Applications (IISA)}}.
  \bibinfo{pages}{1--7}.
\newblock
\urldef\tempurl%
\url{https://doi.org/10.1109/IISA.2019.8900742}
\showDOI{\tempurl}


\bibitem[\protect\citeauthoryear{Ding and Chandra}{Ding and Chandra}{2019}]%
        {ding2019towards}
\bibfield{author}{\bibinfo{person}{Jian Ding} {and} \bibinfo{person}{Ranveer
  Chandra}.} \bibinfo{year}{2019}\natexlab{}.
\newblock \showarticletitle{Towards Low Cost Soil Sensing Using Wi-Fi}. In
  \bibinfo{booktitle}{\emph{The 25th Annual International Conference on Mobile
  Computing and Networking}} (Los Cabos, Mexico)
  \emph{(\bibinfo{series}{MobiCom '19})}. \bibinfo{publisher}{Association for
  Computing Machinery}, \bibinfo{address}{New York, NY, USA}, Article
  \bibinfo{articleno}{39}, \bibinfo{numpages}{16}~pages.
\newblock
\showISBNx{9781450361699}
\urldef\tempurl%
\url{https://doi.org/10.1145/3300061.3345440}
\showDOI{\tempurl}


\bibitem[\protect\citeauthoryear{Dong and Vuran}{Dong and Vuran}{2011}]%
        {dong2011}
\bibfield{author}{\bibinfo{person}{Xin Dong} {and} \bibinfo{person}{Mehmet~C.
  Vuran}.} \bibinfo{year}{2011}\natexlab{}.
\newblock \showarticletitle{A Channel Model for Wireless Underground Sensor
  Networks Using Lateral Waves}. In \bibinfo{booktitle}{\emph{2011 IEEE Global
  Telecommunications Conference - GLOBECOM 2011}}. \bibinfo{pages}{1--6}.
\newblock
\urldef\tempurl%
\url{https://doi.org/10.1109/GLOCOM.2011.6134437}
\showDOI{\tempurl}


\bibitem[\protect\citeauthoryear{Farm}{Farm}{2021}]%
        {autofarm}
\bibfield{author}{\bibinfo{person}{Illinois~Autonomous Farm}.}
  \bibinfo{year}{2021}\natexlab{}.
\newblock \bibinfo{title}{Illinois Autonomous Farm}.
\newblock
  \bibinfo{howpublished}{\url{https://digitalag.illinois.edu/autonomous-farm/}}.
\newblock


\bibitem[\protect\citeauthoryear{Fieldscout}{Fieldscout}{[n.d.]}]%
        {fieldscout}
\bibfield{author}{\bibinfo{person}{Fieldscout}.}
  \bibinfo{year}{[n.d.]}\natexlab{}.
\newblock \bibinfo{title}{FieldScout TDR 350 Soil Moisture Meter}.
\newblock
  \bibinfo{howpublished}{\url{https://www.specmeters.com/soil-and-water/soil-moisture/fieldscout-tdr-meters/fieldscout-tdr-350-soil-moisture-meter-with-case/}}.
\newblock


\bibitem[\protect\citeauthoryear{Fieldscout}{Fieldscout}{2021}]%
        {fieldscoutmanual}
\bibfield{author}{\bibinfo{person}{Fieldscout}.}
  \bibinfo{year}{2021}\natexlab{}.
\newblock \bibinfo{title}{FieldScout TDR 350 Soil Moisture Meter Product
  Manual}.
\newblock
  \bibinfo{howpublished}{\url{https://www.specmeters.com/assets/1/22/6430TDR300_(web).pdf}}.
\newblock


\bibitem[\protect\citeauthoryear{for Digital~Agriculture}{for
  Digital~Agriculture}{2021}]%
        {digitalag}
\bibfield{author}{\bibinfo{person}{Center for Digital~Agriculture}.}
  \bibinfo{year}{2021}\natexlab{}.
\newblock \bibinfo{title}{Center for Digital Agriculture}.
\newblock \bibinfo{howpublished}{\url{https://digitalag.illinois.edu/}}.
\newblock


\bibitem[\protect\citeauthoryear{Gjengset, Xiong, McPhillips, and
  Jamieson}{Gjengset et~al\mbox{.}}{2014}]%
        {gjengset2014}
\bibfield{author}{\bibinfo{person}{Jon Gjengset}, \bibinfo{person}{Jie Xiong},
  \bibinfo{person}{Graeme McPhillips}, {and} \bibinfo{person}{Kyle Jamieson}.}
  \bibinfo{year}{2014}\natexlab{}.
\newblock \showarticletitle{Phaser: Enabling Phased Array Signal Processing on
  Commodity WiFi Access Points}. In \bibinfo{booktitle}{\emph{Proceedings of
  the 20th Annual International Conference on Mobile Computing and Networking}}
  (Maui, Hawaii, USA) \emph{(\bibinfo{series}{MobiCom '14})}.
  \bibinfo{publisher}{Association for Computing Machinery},
  \bibinfo{address}{New York, NY, USA}, \bibinfo{pages}{153–164}.
\newblock
\showISBNx{9781450327831}
\urldef\tempurl%
\url{https://doi.org/10.1145/2639108.2639139}
\showDOI{\tempurl}


\bibitem[\protect\citeauthoryear{Godfray, Beddington, Crute, Haddad, Lawrence,
  Muir, Pretty, Robinson, Thomas, and Toulmin}{Godfray et~al\mbox{.}}{2010}]%
        {Godfray812}
\bibfield{author}{\bibinfo{person}{H.~Charles~J. Godfray},
  \bibinfo{person}{John~R. Beddington}, \bibinfo{person}{Ian~R. Crute},
  \bibinfo{person}{Lawrence Haddad}, \bibinfo{person}{David Lawrence},
  \bibinfo{person}{James~F. Muir}, \bibinfo{person}{Jules Pretty},
  \bibinfo{person}{Sherman Robinson}, \bibinfo{person}{Sandy~M. Thomas}, {and}
  \bibinfo{person}{Camilla Toulmin}.} \bibinfo{year}{2010}\natexlab{}.
\newblock \showarticletitle{Food Security: The Challenge of Feeding 9 Billion
  People}.
\newblock \bibinfo{journal}{\emph{Science}} \bibinfo{volume}{327},
  \bibinfo{number}{5967} (\bibinfo{year}{2010}), \bibinfo{pages}{812--818}.
\newblock
\showISSN{0036-8075}
\urldef\tempurl%
\url{https://doi.org/10.1126/science.1185383}
\showDOI{\tempurl}
\showeprint{https://science.sciencemag.org/content/327/5967/812.full.pdf}


\bibitem[\protect\citeauthoryear{Gro}{Gro}{2021}]%
        {miraclegro}
\bibfield{author}{\bibinfo{person}{Miracle Gro}.}
  \bibinfo{year}{2021}\natexlab{}.
\newblock \bibinfo{title}{Miracle-Gro® Potting Mix}.
\newblock
  \bibinfo{howpublished}{\url{https://www.miraclegro.com/en-us/products/soils/miracle-gro-potting-mix}}.
\newblock


\bibitem[\protect\citeauthoryear{H{\"o}nig and Kamra}{H{\"o}nig and
  Kamra}{2016}]%
        {wolfgang2016L}
\bibfield{author}{\bibinfo{person}{Wolfgang H{\"o}nig} {and}
  \bibinfo{person}{Nitin Kamra}.} \bibinfo{year}{2016}\natexlab{}.
\newblock \showarticletitle{RF-Based Relative Localization for Robot Swarms}.
\newblock


\bibitem[\protect\citeauthoryear{Islam, Islam, Kaur, and Nirjon}{Islam
  et~al\mbox{.}}{2019}]%
        {islam2019lorain}
\bibfield{author}{\bibinfo{person}{Bashima Islam}, \bibinfo{person}{Md~Tamzeed
  Islam}, \bibinfo{person}{Jasleen Kaur}, {and} \bibinfo{person}{Shahriar
  Nirjon}.} \bibinfo{year}{2019}\natexlab{}.
\newblock \showarticletitle{Lorain: Making a case for lora in indoor
  localization}. In \bibinfo{booktitle}{\emph{2019 IEEE International
  Conference on Pervasive Computing and Communications Workshops (PerCom
  Workshops)}}. IEEE, \bibinfo{pages}{423--426}.
\newblock


\bibitem[\protect\citeauthoryear{Josephson, Barnhart, Katti, Winstein, and
  Chandra}{Josephson et~al\mbox{.}}{2019}]%
        {josephson2019}
\bibfield{author}{\bibinfo{person}{Colleen Josephson}, \bibinfo{person}{Bradley
  Barnhart}, \bibinfo{person}{Sachin Katti}, \bibinfo{person}{Keith Winstein},
  {and} \bibinfo{person}{Ranveer Chandra}.} \bibinfo{year}{2019}\natexlab{}.
\newblock \bibinfo{title}{RF Soil Moisture Sensing via Radar Backscatter Tags}.
\newblock
\newblock
\showeprint[arxiv]{1912.12382}~[eess.SP]


\bibitem[\protect\citeauthoryear{Kotaru, Joshi, Bharadia, and Katti}{Kotaru
  et~al\mbox{.}}{2015}]%
        {spotfi}
\bibfield{author}{\bibinfo{person}{Manikanta Kotaru}, \bibinfo{person}{Kiran
  Joshi}, \bibinfo{person}{Dinesh Bharadia}, {and} \bibinfo{person}{Sachin
  Katti}.} \bibinfo{year}{2015}\natexlab{}.
\newblock \showarticletitle{SpotFi: Decimeter Level Localization Using WiFi}.
  In \bibinfo{booktitle}{\emph{Proceedings of the 2015 ACM Conference on
  Special Interest Group on Data Communication}} (London, United Kingdom)
  \emph{(\bibinfo{series}{SIGCOMM '15})}. \bibinfo{publisher}{Association for
  Computing Machinery}, \bibinfo{address}{New York, NY, USA},
  \bibinfo{pages}{269–282}.
\newblock
\showISBNx{9781450335423}
\urldef\tempurl%
\url{https://doi.org/10.1145/2785956.2787487}
\showDOI{\tempurl}


\bibitem[\protect\citeauthoryear{Lin, Hao, Yu, Zheng, and He}{Lin
  et~al\mbox{.}}{2019}]%
        {lin2019underground}
\bibfield{author}{\bibinfo{person}{Kaiqiang Lin}, \bibinfo{person}{Tong Hao},
  \bibinfo{person}{Zhouwei Yu}, \bibinfo{person}{Wuan Zheng}, {and}
  \bibinfo{person}{Wenchao He}.} \bibinfo{year}{2019}\natexlab{}.
\newblock \showarticletitle{A Preliminary Study of UG2AG Link Quality in
  LoRa-based Wireless Underground Sensor Networks}. In
  \bibinfo{booktitle}{\emph{2019 IEEE 44th Conference on Local Computer
  Networks (LCN)}}. \bibinfo{pages}{51--59}.
\newblock
\urldef\tempurl%
\url{https://doi.org/10.1109/LCN44214.2019.8990756}
\showDOI{\tempurl}


\bibitem[\protect\citeauthoryear{Pedregosa, Varoquaux, Gramfort, Michel,
  Thirion, Grisel, Blondel, Prettenhofer, Weiss, Dubourg, Vanderplas, Passos,
  Cournapeau, Brucher, Perrot, and Duchesnay}{Pedregosa et~al\mbox{.}}{2011}]%
        {scikit-learn}
\bibfield{author}{\bibinfo{person}{F. Pedregosa}, \bibinfo{person}{G.
  Varoquaux}, \bibinfo{person}{A. Gramfort}, \bibinfo{person}{V. Michel},
  \bibinfo{person}{B. Thirion}, \bibinfo{person}{O. Grisel},
  \bibinfo{person}{M. Blondel}, \bibinfo{person}{P. Prettenhofer},
  \bibinfo{person}{R. Weiss}, \bibinfo{person}{V. Dubourg}, \bibinfo{person}{J.
  Vanderplas}, \bibinfo{person}{A. Passos}, \bibinfo{person}{D. Cournapeau},
  \bibinfo{person}{M. Brucher}, \bibinfo{person}{M. Perrot}, {and}
  \bibinfo{person}{E. Duchesnay}.} \bibinfo{year}{2011}\natexlab{}.
\newblock \showarticletitle{Scikit-learn: Machine Learning in {P}ython}.
\newblock \bibinfo{journal}{\emph{Journal of Machine Learning Research}}
  \bibinfo{volume}{12} (\bibinfo{year}{2011}), \bibinfo{pages}{2825--2830}.
\newblock


\bibitem[\protect\citeauthoryear{Vasisht, Kapetanovic, Won, Jin, Chandra,
  Sinha, Kapoor, Sudarshan, and Stratman}{Vasisht et~al\mbox{.}}{2017}]%
        {vasisht2017farmbeats}
\bibfield{author}{\bibinfo{person}{Deepak Vasisht}, \bibinfo{person}{Zerina
  Kapetanovic}, \bibinfo{person}{Jongho Won}, \bibinfo{person}{Xinxin Jin},
  \bibinfo{person}{Ranveer Chandra}, \bibinfo{person}{Sudipta Sinha},
  \bibinfo{person}{Ashish Kapoor}, \bibinfo{person}{Madhusudhan Sudarshan},
  {and} \bibinfo{person}{Sean Stratman}.} \bibinfo{year}{2017}\natexlab{}.
\newblock \showarticletitle{FarmBeats: An IoT Platform for Data-Driven
  Agriculture}. In \bibinfo{booktitle}{\emph{14th {USENIX} Symposium on
  Networked Systems Design and Implementation ({NSDI} 17)}}.
  \bibinfo{publisher}{{USENIX} Association}, \bibinfo{address}{Boston, MA},
  \bibinfo{pages}{515--529}.
\newblock
\showISBNx{978-1-931971-37-9}
\urldef\tempurl%
\url{https://www.usenix.org/conference/nsdi17/technical-sessions/presentation/vasisht}
\showURL{%
\tempurl}


\bibitem[\protect\citeauthoryear{Vasisht, Kumar, and Katabi}{Vasisht
  et~al\mbox{.}}{2016}]%
        {chronos}
\bibfield{author}{\bibinfo{person}{Deepak Vasisht}, \bibinfo{person}{Swarun
  Kumar}, {and} \bibinfo{person}{Dina Katabi}.}
  \bibinfo{year}{2016}\natexlab{}.
\newblock \showarticletitle{Decimeter-Level Localization with a Single WiFi
  Access Point}. In \bibinfo{booktitle}{\emph{13th {USENIX} Symposium on
  Networked Systems Design and Implementation ({NSDI} 16)}}.
  \bibinfo{publisher}{{USENIX} Association}, \bibinfo{address}{Santa Clara,
  CA}, \bibinfo{pages}{165--178}.
\newblock
\showISBNx{978-1-931971-29-4}
\urldef\tempurl%
\url{https://www.usenix.org/conference/nsdi16/technical-sessions/presentation/vasisht}
\showURL{%
\tempurl}


\bibitem[\protect\citeauthoryear{Vuran and Silva}{Vuran and Silva}{2009}]%
        {vuran2009}
\bibfield{author}{\bibinfo{person}{M Vuran} {and} \bibinfo{person}{Agnelo
  Silva}.} \bibinfo{year}{2009}\natexlab{}.
\newblock \bibinfo{booktitle}{\emph{Communication Through Soil in Wireless
  Underground Sensor Networks – Theory and Practice}}.
\newblock \bibinfo{pages}{309--347}.
\newblock
\showISBNx{978-3-642-01340-9}
\urldef\tempurl%
\url{https://doi.org/10.1007/978-3-642-01341-6_12}
\showDOI{\tempurl}


\bibitem[\protect\citeauthoryear{Wan, Yang, Cui, and Sardar}{Wan
  et~al\mbox{.}}{2017}]%
        {wan2017lora}
\bibfield{author}{\bibinfo{person}{Xue-fen Wan}, \bibinfo{person}{Yi Yang},
  \bibinfo{person}{Jian Cui}, {and} \bibinfo{person}{Muhammad~Sohail Sardar}.}
  \bibinfo{year}{2017}\natexlab{}.
\newblock \showarticletitle{Lora propagation testing in soil for wireless
  underground sensor networks}. In \bibinfo{booktitle}{\emph{2017 Sixth
  Asia-Pacific Conference on Antennas and Propagation (APCAP)}}.
  \bibinfo{pages}{1--3}.
\newblock
\urldef\tempurl%
\url{https://doi.org/10.1109/APCAP.2017.8420657}
\showDOI{\tempurl}


\bibitem[\protect\citeauthoryear{Wang, Chang, Aggarwal, Abari, and Keshav}{Wang
  et~al\mbox{.}}{2020}]%
        {wang2020soil}
\bibfield{author}{\bibinfo{person}{Ju Wang}, \bibinfo{person}{Liqiong Chang},
  \bibinfo{person}{Shourya Aggarwal}, \bibinfo{person}{Omid Abari}, {and}
  \bibinfo{person}{Srinivasan Keshav}.} \bibinfo{year}{2020}\natexlab{}.
\newblock \showarticletitle{Soil Moisture Sensing with Commodity RFID Systems}.
  In \bibinfo{booktitle}{\emph{Proceedings of the 18th International Conference
  on Mobile Systems, Applications, and Services}} (Toronto, Ontario, Canada)
  \emph{(\bibinfo{series}{MobiSys '20})}. \bibinfo{publisher}{Association for
  Computing Machinery}, \bibinfo{address}{New York, NY, USA},
  \bibinfo{pages}{273–285}.
\newblock
\showISBNx{9781450379540}
\urldef\tempurl%
\url{https://doi.org/10.1145/3386901.3388940}
\showDOI{\tempurl}


\bibitem[\protect\citeauthoryear{Wijeratne, Kiv, Aker, Talebi, and
  Lary}{Wijeratne et~al\mbox{.}}{2020}]%
        {wijeratne2020}
\bibfield{author}{\bibinfo{person}{Lakitha~O.H. Wijeratne},
  \bibinfo{person}{Daniel~R. Kiv}, \bibinfo{person}{Adam~R. Aker},
  \bibinfo{person}{Shawhin Talebi}, {and} \bibinfo{person}{David~J. Lary}.}
  \bibinfo{year}{2020}\natexlab{}.
\newblock \showarticletitle{Using Machine Learning for the Calibration of
  Airborne Particulate Sensors}.
\newblock \bibinfo{journal}{\emph{Sensors}} \bibinfo{volume}{20},
  \bibinfo{number}{1} (\bibinfo{year}{2020}).
\newblock
\showISSN{1424-8220}
\urldef\tempurl%
\url{https://doi.org/10.3390/s20010099}
\showDOI{\tempurl}


\end{thebibliography}

\end{document}